\documentclass[twocolumn,amsmath,amssymb,prl,superscriptaddress,a4paper,floatfix,showkeys]{revtex4}
\usepackage{graphicx}
\usepackage{mathptmx}      

\begin{document}

\title{Corpuscular Event-by-Event Simulation of Quantum Optics Experiments\footnote{
Invited paper presented at FQMT11.\\
Accepted for publication in Physica Scripta 27 June 2012.}:\\
Application to a Quantum-Controlled Delayed-Choice Experiment}

\author{Hans De Raedt}
\email{h.a.de.raedt@rug.nl}
\affiliation{%
Department of Applied Physics,
Zernike Institute for Advanced Materials,
University of Groningen, Nijenborgh 4, NL-9747 AG Groningen, The Netherlands
}%
\author{M. Delina}
\email{m.m.delina@rug.nl}
\affiliation{%
Department of Applied Physics,
Zernike Institute for Advanced Materials,
University of Groningen, Nijenborgh 4, NL-9747 AG Groningen, The Netherlands
}%
\affiliation{%
Physics Department,
Faculty of Mathematics and Natural Science,
State University of Jakarta,
Jl.Permuda No 10, Jakarta 13220, Indonesia
}%
\author{Fengping Jin}
\email{f.jin@fz-juelich.de}           
\affiliation{%
Institute for Advanced Simulation, J\"ulich Supercomputing Centre,
Forschungzentrum J\"ulich, D-52425 J\"ulich, Germany
}%
\author{Kristel Michielsen}
\email{k.michielsen@fz-juelich.de}           
\affiliation{%
Institute for Advanced Simulation, J\"ulich Supercomputing Centre,
Forschungzentrum J\"ulich, D-52425 J\"ulich, Germany
}%
\affiliation{%
RWTH Aachen University, D-52056 Aachen, Germany
}%

\begin{abstract}
A corpuscular simulation model of optical phenomena that does not require
the knowledge of the solution of a wave equation of the whole system
and reproduces the results of Maxwell's theory by generating detection events one-by-one is discussed.
The event-based corpuscular model gives a unified description of multiple-beam fringes
of a plane parallel plate and single-photon Mach-Zehnder interferometer,
Wheeler's delayed choice, photon tunneling, quantum eraser,
two-beam interference, Einstein-Podolsky-Rosen-Bohm
and Hanbury Brown-Twiss experiments.
The approach is illustrated by application to a recent proposal 
for a quantum-controlled delayed choice experiment,
demonstrating that also this thought experiment can be
understood in terms of particle processes only.
\end{abstract}

\keywords{Interference, quantum theory, discrete-event simulation}
\date{\today}

\maketitle

\section{Introduction}

Quantum theory has proven extraordinarily powerful for describing the
statistical properties of a vast number of laboratory experiments.
Conceptually, it is straightforward to use the quantum theoretical formalism to
calculate numbers that can be compared with experimental data, at least if
these numbers refer to statistical averages.
However, a fundamental problem appears if an experiment provides
access to the individual events that collectively build the statistical average.
Prime examples are the single-electron two-slit experiment~\cite{TONO98},
neutron interferometry experiments~\cite{RAUC00} and
similar experiments in optics where the click
of the detector is identified with the arrival of a single photon~\cite{GARR09}.
Although quantum theory provides a recipe to compute the frequencies for observing events
it does not account for the observation of the individual detection events themselves~\cite{HOME97,BALL03}.
For a recent review of various approaches to the quantum measurement problem and an explanation of it within
the statistical interpretation, see Ref.~\onlinecite{NIEU11b}.

From the viewpoint of quantum theory, the central issue is how it can be that
experiments yield definite answers.
As stated by Leggett~\cite{LEGG86}: ``In the final analysis,
physics cannot forever refuse to give an account of how it is
that we obtain definite results whenever we do a particular
measurement''.

This paper is not about interpretations or extensions of quantum theory.
It gives a brief account of a very different approach to
deal with the fact that experiments yield definite results.
The latter, which is intimately linked to human perception,
is taken as fundamental. We call these definite results ``events''.
Instead of trying to fit the existence of these events
in some formal, mathematical theory, we change the paradigm
by directly searching for the rules that transform events
into other events and, by repeated application,
yield frequency distributions of events that
agree with those predicted by quantum theory.
Obviously, such rules cannot be derived from quantum theory or,
as a matter of fact, of any theory that is probabilistic in nature
simply because these theories do not entail a procedure (= algorithm)
to produce events themselves.

The event-based approach has successfully been
used to perform discrete-event simulations of the single beam splitter and
Mach-Zehnder interferometer experiment
of Grangier {\sl et al.}~\cite{GRAN86} (see Refs.~\cite{RAED05d,RAED05b,MICH11a}),
Wheeler's delayed choice experiment of Jacques {\sl et al.}~\cite{JACQ07}
(see Refs.~\cite{ZHAO08b,MICH10a,MICH11a}),
the quantum eraser experiment of Schwindt {\sl et al.}~\cite{SCHW99} (see Ref.~\cite{JIN10c,MICH11a}),
double-slit and two-beam single-photon interference experiments and the single-photon interference experiment with
a Fresnel biprism of Jacques {\sl et al.}~\cite{JACQ05} (see Ref.~\cite{JIN10b,MICH11a}),
quantum cryptography protocols (see Ref.~\cite{ZHAO08a}),
the Hanbury Brown-Twiss experiment of Agafonov {\sl et al.}~\cite{AGAF08} (see Ref.~\cite{JIN10a,MICH11a}),
universal quantum computation (see Ref.~\cite{RAED05c,MICH05}),
Einstein-Podolsky-Rosen-Bohm-type of experiments of Aspect {\sl et al.}~\cite{ASPE82a,ASPE82b}
and Weihs {\sl et al.}~\cite{WEIH98} (see Refs.~\cite{RAED06c,RAED07a,RAED07b,RAED07c,RAED07d,ZHAO08,MICH11a}),
and the propagation of electromagnetic plane waves through homogeneous thin films and stratified media (see Ref.~\cite{TRIE11,MICH11a}).
An extensive review of the simulation method and its applications is given in Ref.~\cite{MICH11a}.

A detailed discussion of the discrete-event approach cannot be fitted in this short paper.
Therefore, we have chosen to illustrate the approach by an application
to a recent proposal for a quantum-controlled Wheeler delayed choice experiment~\cite{IONI11}.
We demonstrate that also this thought experiment can be
understood in terms of event-based, particle-like processes only.
The presentation is sufficiently detailed such that the reader
who is interested can reproduce our results.

\section{Wheeler's delayed-choice experiment}

\begin{figure}[t]
\begin{center}
\includegraphics[width=8cm ]{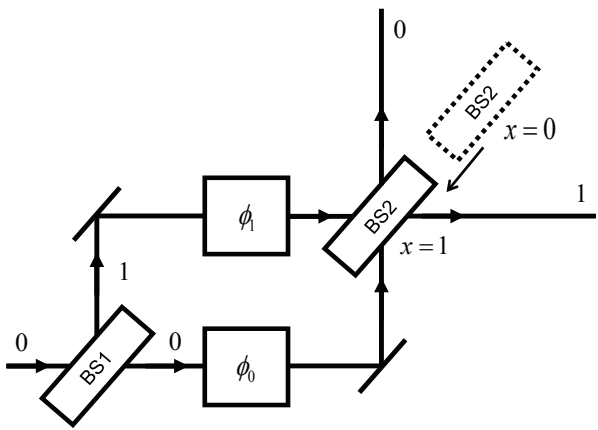}
\caption{%
Diagram of a standard Wheeler delayed-choice experiment with a Mach-Zehnder interferometer.
Photons enter the interferometer via 50--50 beam splitter 1 (BS1).
In the wave picture, the partial wave traveling along path 0 (1)
acquires a phase shift $\phi_0$ ($\phi_1$).
The variable $x=0,1$ controls the presence of 50--50 beam splitter 2 (BS2).
If BS2 is not in place ($x=0$, indicated by the dashed rectangle)
the partial waves do not interfere
and the probability to observe the photon in path $0$ or $1$
does not depend on the phase shifts.
If BS2 is in place ($x=1$, indicated by solid rectangle)
the partial waves interfere
and the probability to observe the photon in path $0$ or $1$
is given by $(1+\cos(\phi_0-\phi_1))/2$ or $(1-\cos(\phi_0-\phi_1))/2$,
respectively.
}%
\label{fig.1}
\end{center}
\end{figure}

Particle-wave duality, a concept of quantum theory, attributes to photons the
properties of both wave and particle behavior depending upon the circumstances of
the experiment~\cite{HOME97}.
The particle behavior of photons has been shown in an experiment composed of a
single beam splitter (BS) and a source emitting single photons and pairs of
photons~\cite{GRAN86}. The wave character has been demonstrated in a single-photon Mach-Zehnder
interferometer (MZI) experiment~\cite{GRAN86}.
The layout of such an experiment is shown in Fig.~\ref{fig.1}.
By adding a device which controls the presence or absence of the second beam splitter BS2,
this setup can be used to perform a delayed-choice experiment.
Originally, Wheeler proposed a double-slit gedanken
experiment in which the decision to observe wave or particle behavior is made after the photon has
passed the slits~\cite{WHEE83}.
Similarly, in the MZI experiment, the decision to remove and place BS2 at the intersection
of paths 0 and 1 can, in principle, be made after the photon has passed BS1.
The conclusion is that the pictorial description of this experiment defies common sense:
The behavior of the photon in the past is said to be changing from a particle to a wave or vice versa.

\begin{figure}[t]
\begin{center}
\includegraphics[width=8cm ]{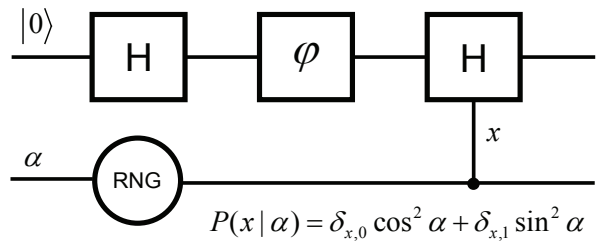}
\caption{%
Quantum gate representation of the standard Wheeler delayed-choice
experiment with a Mach-Zehnder interferometer (see Fig.~\ref{fig.1}).
The first Hadamard gate ${\bf H}$ acts as a 50-50 beam splitter,
changing the state $|0\rangle$ into the state $(|0\rangle+|1\rangle)/sqrt{2}$.
The phase gate $\varphi$ changes the amplitude
of the state $|1\rangle$ by $e^{i\varphi}$.
The second (controlled) Hadamard gate ${\bf H}$ act as a 50-50 beam splitter
if the control variable $x=1$ or passes the photons unaltered if $x=0$.
The angle $\alpha$ determines the probability
that the control variable $x$ is 1.
A pair of detectors (not shown) signals the presence
of a photon in the state $|0\rangle$ or $|1\rangle$
and with each detected photon the value of $x$ is being recorded.
}%
\label{fig.2}
\end{center}
\end{figure}

\begin{figure}[t]
\begin{center}
\includegraphics[width=8cm ]{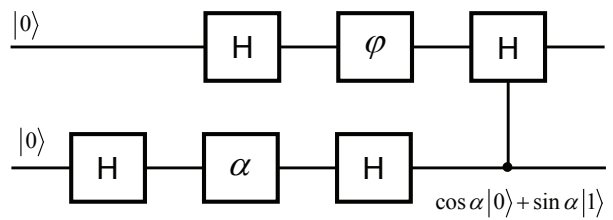}
\caption{%
Quantum gate representation of the quantum version of Wheeler delayed-choice
experiment with a Mach-Zehnder interferometer~\cite{IONI11}.
Reading from left to right,
the first Hadamard gate ${\bf H}$ on the top line acts as a 50-50 beam splitter
and the phase gate $\varphi$ changes the amplitude
of the state $|1\rangle$ by $e^{i\varphi}$.
The second (controlled) Hadamard gate ${\bf H}$ on the top line acts as a 50-50 beam splitter
if the state of the ancilla is $|1\rangle$
or passes the photons unaltered if that state is $|0\rangle$.
Initially in the state $|0\rangle$,
the ancilla is prepared in a uniform superposition
of the states $|0\rangle$ and $|1\rangle$
by another interferometer circuit (bottom line)
in which the phase gate $\alpha$ changes the amplitude
of the ancilla state $|1\rangle$ by $e^{i\alpha}$.
The angle $\alpha$ determines the probabilities
of the states $|0\rangle$ and $|1\rangle$.
A pair of detectors (not shown) signals the presence
of the photon in the state $|0\rangle$ or $|1\rangle$.
Similarly, another pair of detectors (not shown) signals the presence
of the ancilla in the state $|0\rangle$ or $|1\rangle$.
}%
\label{fig.3}
\end{center}
\end{figure}

\section{Quantum controlled delayed-choice experiment}

It is of interest to enquire what happens if the
variable $x$ which controls the presence of BS2 (see Fig.~\ref{fig.1})
or, equivalently, the controlled Hadamard gate (see Fig.~\ref{fig.2})
is replaced by a quantum two-state system~\cite{IONI11}.
In a sense, one could then view the experiment as a simple example
of a quantum-controlled experiment~\cite{IONI11}.
The original proposal of the quantum-controlled delayed-choice experiment~\cite{IONI11}
is formulated in a notation that is commonly used in the
quantum computer literature~\cite{NIEL00}.
To facilitate the comparison with this work, we
also adopt to this notation from now on.
First, in Fig.~\ref{fig.2} we show the quantum gate diagram
that is equivalent to the standard delayed-choice experiment
depicted in Fig.~\ref{fig.1}.
The main change, irrelevant from a conceptual point of view,
is to replace the beam splitters by Hadamard gates.
In Ref.~\onlinecite{IONI11}, it is proposed to replace the
classical random variable $x$ in Fig.~\ref{fig.2}
by a qubit, conventionally called ancilla,
that can be in a superposition of the states
$|0\rangle$ and $|1\rangle$.
As shown in Fig.~\ref{fig.3}, the state of the ancilla
controls the operation of the last Hadamard gate on the top line.
In our implementation, we have chosen to include a
preparation procedure for the state of the ancilla,
as indicated in Fig.~\ref{fig.3}.

For completeness and comparison with the event-by-event simulation data,
we give the quantum-theoretical description of this experiment
in terms of the state $|vu\rangle=|v\rangle\otimes |u\rangle$
where $u,v=0,1$ label the basis states and $|u\rangle$
and $|v\rangle$ denote the state of the ancilla and photon,
respectively.
The amplitudes at the input
${\mathbf a}=\left(a_{00},a_{01},a_{10},a_{11}\right)^T$
and output
${\mathbf b}=\left(b_{00},b_{01},b_{10},b_{11}\right)^T$
of the experiment
depicted in Fig.~\ref{fig.3} are related by
\begin{eqnarray}
{\mathbf b}
&=&
\left(\begin{array}{cccc}
        \phantom{-}1 & \phantom{-}0 & \phantom{-}a &   \phantom{-}0 \\
        \phantom{-}0 & \phantom{-}a & \phantom{-}0 &   \phantom{-}a \\
        \phantom{-}1 & \phantom{-}0 & \phantom{-}1 &   \phantom{-}0 \\
        \phantom{-}0 &           -a & \phantom{-}0 &   \phantom{-}a
\end{array}\right)
\left(\begin{array}{cccc}
        1 & 0 & 0 & 0 \\
        0 & 1 & 0 & 0 \\
        0 & 0& 1 & 0 \\
        0 & 0& 0 & e^{i\varphi}
\end{array}\right)
\nonumber \\
&\times&
\left(\begin{array}{cccc}
        \phantom{-}a & \phantom{-}0 & \phantom{-}a &   \phantom{-}0 \\
        \phantom{-}0 & \phantom{-}a & \phantom{-}0 &   \phantom{-}a \\
                  -a & \phantom{-}0 & \phantom{-}a &   \phantom{-}0 \\
        \phantom{-}0 &           -a & \phantom{-}0 &   \phantom{-}a
\end{array}\right)
\left(\begin{array}{cccc}
        \phantom{-}a & \phantom{-}a &  \phantom{-}0 &  \phantom{-}0 \\
                  -a & \phantom{-}a &  \phantom{-}0 &  \phantom{-}0 \\
        \phantom{-}0 & \phantom{-}0 &  \phantom{-}a &  \phantom{-}a \\
        \phantom{-}0 & \phantom{-}0 &            -a &  \phantom{-}a
\end{array}\right)
\nonumber \\
&\times&
\left(\begin{array}{cccc}
        1 & 0 & 0 & 0 \\
        0 & e^{i\alpha}  & 0 & 0 \\
        0 & 0& 1 & 0 \\
        0 & 0& 0 & 1
\end{array}\right)
\left(\begin{array}{cccc}
        \phantom{-}a & \phantom{-}a &  \phantom{-}0 &  \phantom{-}0 \\
                  -a & \phantom{-}a &  \phantom{-}0 &  \phantom{-}0 \\
        \phantom{-}0 & \phantom{-}0 &  \phantom{-}a &  \phantom{-}a \\
        \phantom{-}0 & \phantom{-}0 &            -a &  \phantom{-}a
\end{array}\right)
{\mathbf a}
,
\label{app14}
\end{eqnarray}
where $a=1/\sqrt{2}$.

Reading from right to left, the matrices in Eq.~(\ref{app14}) represent the action of a Hadamard operation on the ancilla,
a phase shift (by $\alpha$) operation on the ancilla,
another Hadamard operation on the ancilla,
a Hadamard operation on the photon,
a phase shift (by $\varphi$) operation on the photon,
and a controlled (by the ancilla) Hadamard operation on the photon.
Note that all these operations only affect the state, that is the wave function,
which describes the statistical properties of the whole system and cannot be
interpreted as having causal effects on a particular particle
without running into conceptual and logical problems~\cite{HOME97}.

For the case at hand, $a_{00}=1$ and all other $a$'s are zero.
Then it follows from Eq.~(\ref{app14}) that the probability to detect a pair (photon,ancilla)
in the state $|vu\rangle$ is given by $p(v,u)=|b_{v,u}|^2$.
More explicitly we have~\cite{IONI11}
\begin{eqnarray}
p(v=0,u=0)&=&\frac{1}{2}\cos^2\alpha ,
\nonumber \\
p(v=1,u=0)&=&\frac{1}{2}\cos^2\alpha ,
\nonumber \\
p(v=0,u=1)&=&\sin^2\alpha\cos^2\frac{\varphi}{2},
\nonumber \\
p(v=1,u=1)&=&\sin^2\alpha\sin^2\frac{\varphi}{2}
.
\label{app15}
\end{eqnarray}
Note that Eq.~(\ref{app15}) is identical to the corresponding
result for the standard delayed-choice experiment.

\section{Simulation model}\label{sec2}

The model presented in this paper
builds on earlier work~\cite{RAED05d,RAED05b,RAED05c,MICH05,RAED06c,RAED07a,RAED07b,RAED07c,ZHAO08,MICH11a} in which we have
demonstrated that it may be possible to simulate quantum phenomena on the level of individual events
without invoking concepts of quantum theory.

In our simulation approach, a messenger (representing the photon or the ancilla),
carries a message (representing the phase)
and is routed through the network and the various units
that process the messages.

We now explicitly describe our simulation model that
is, we specify the message carried by the messengers, the algorithms that
simulate the processing units and the data analysis procedure.

\textbf{Messenger}. Particles carry a message represented by a
two-dimensional unit vector ${\mathbf y}_{k,n}=\left( \cos \psi _{k,n},\sin \psi _{k,n}\right) $ where $\psi _{k,n}$ refers to the phase of
the photon. The subscript $n\geq 0$ numbers the consecutive messages and $k=0,1$ labels the port of the beam splitter at which the message arrives.
Every time, a messenger is created, the message is initialized to ${\mathbf y}_{k,n}=(1,0)$.

\textbf{Hadamard gate}. The key element of the event-by-event
approach is a processing unit that is adaptive, that is it can learn
from the messengers that arrive at its input ports~\cite{RAED05d,RAED05b,MICH11a}.
The processing unit consists of an input stage called deterministic learning machine (DLM)~\cite{RAED05d,RAED05b},
a transformation stage, and an output stage.
In experiments with single particles,
the input stage receives a message on either input port $k=0$ or $k=1$,
but never on both ports simultaneously.
The arrival of a message on port 0 (1) corresponds to an event of type 0 (1).
The input events are represented by the vectors ${\bf e}_{n}=(1,0)$ or ${\bf e}_{n}=(0,1)$ if the $n$th
event occurred on port 0 or 1, respectively. The DLM has two sets of internal
registers $( C_{k,n},S_{k,n})$ and one
internal vector ${\bf x}_{n}=( x_{0,n},x_{1,n}) $,
where $x_{0,n}+x_{1,n}=1$ and $x_{i,n}>0$.
These three two-dimensional vectors are labeled by the message number $n$ because their
content is updated every time the DLM receives a message.
Thus, the DLM can only store 6 numbers, not more.
Before the simulation starts we set ${\bf x}_{0}=(x_{0,0},x_{1,0}) =( {\cal R},1-{\cal R}) $,
where ${\cal R}$ is a uniform pseudo-random number.
In a similar way, we use pseudo-random numbers to set $( C_{k,0},S_{k,0})$ for $k=0,1$.
Upon receiving the $(n+1)$th input event, the DLM performs the following steps:
(1) it stores the message ${\mathbf y}_{k,n+1}=( \cos \psi_{k,n+1},\sin \psi _{k,n+1}) $
in its internal register $( C_{k,n+1},S_{k,n+1})$ and (2) it updates its internal vector according to the rule
\begin{equation}
x_{i,n+1}=\gamma x_{i,n}+( 1-\gamma ) \delta _{i,k},
\end{equation}
where $0<\gamma <1$ is a parameter that controls the learning process. By
construction $x_{0,n+1}+x_{1,n+1}=1$ and $x_{i,n+1}\geq 0$.

The parameter $\gamma$ affects the time that the machine needs to
adapt to a new situation, that is when the ratio of particles on paths 0 and 1 changes.
By reducing $\gamma$, the time to adapt decreases but the accuracy with which the machine
reproduces the ratio also decreases. In the limit that $\gamma=0$, the machine learns nothing:
it simply echoes the last message that it received~\cite{RAED05b,RAED05d}.
If $\gamma\rightarrow1^-$, the machine learns slowly and accurately reproduces the ratio
of particles that enter via path 0 and 1.
It is in this case that the machine can be used to reproduce, event-by-event, the
interference patterns that are characteristic of quantum phenomena~\cite{RAED05b,RAED05d,MICH11a}.

The transformation stage implements the specific functionality
of the unit, the Hadamard operation for the case at hand.
It takes as input the data stored in the two internal registers
$( C_{k,n+1},S_{k,n+1})$ ($k=0,1$)
and in the internal vector ${\bf x}_{n+1}=( x_{0,n+1},x_{1,n+1}) $
and constructs the four-dimensional vector
\begin{equation}
{\mathbf V}=\frac{1}{\sqrt{2}}
\left(
\begin{array}{c}
C_{1,n+1}\sqrt{x_{1,n+1}}+C_{0,n+1}\sqrt{x_{0,n+1}} \\
S_{1,n+1}\sqrt{x_{1,n+1}}+S_{0,n+1}\sqrt{x_{0,n+1}} \\
C_{1,n+1}\sqrt{x_{1,n+1}}-C_{0,n+1}\sqrt{x_{0,n+1}} \\
S_{1,n+1}\sqrt{x_{1,n+1}}-S_{0,n+1}\sqrt{x_{0,n+1}}%
\end{array}%
\right) .
\label{T1}
\end{equation}%
Rewriting this vector as a two-dimensional vector
with complex-valued entries, it is easy to show
that ${\bf V}$ corresponds to the matrix-vector multiplication in the quantum
theoretical description of the Hadamard gate~\cite{MICH05}.

The vector ${\mathbf V}$ is then passed to the output stage
which determines the output port through which the messenger leaves the unit.
The output stage sends the message
\begin{equation}
{\mathbf w}_{0,n+1}=( {\mathbf V}_0+{\mathbf V}_1) /({\mathbf V}_0^{2}+{\mathbf V}_1^{2})^{1/2},
\end{equation}
through output port 0 if
${\mathbf w}_{0,n+1}^{2}<{\cal R}$ where $0<{\cal R}<1$ is
a uniform pseudo-random number.
Otherwise, the output stage sends the message
\begin{equation}
{\mathbf w}_{1,n+1}=( {\mathbf V}_2+{\mathbf V}_3) /({\mathbf V}_2^{2}+{\mathbf V}_3^{2})^{1/2},
\end{equation}
through output port 1.

\textbf{Controlled Hadamard gate}.
The event-based processor of this device is identical
to the one of the Hadamard gate itself except that
the vector ${\mathbf V}$ is computed according to Eq.~(\ref{T1})
if the control bit (called $x$) is 1 only.
If the control bit is 0, ${\mathbf V}$ is given by
\begin{equation}
{\mathbf V}=
\left(
\begin{array}{c}
C_{0,n+1}\sqrt{x_{0,n+1}} \\
S_{0,n+1}\sqrt{x_{0,n+1}} \\
C_{1,n+1}\sqrt{x_{1,n+1}} \\
S_{1,n+1}\sqrt{x_{1,n+1}}%
\end{array}%
\right) .
\label{T2}
\end{equation}%

\textbf{Phase gate}. The unit that performs the phase shift by
an angle $\phi$ changes the message
${\mathbf y}_{k,n}$ according to the rule
\begin{eqnarray}
{\mathbf y}_{0,n}&\leftarrow&
\left(\begin{array}{cc}
        1 & 0 \\
        0 & 1
\end{array}\right)
,
\nonumber \\
{\mathbf y}_{1,n}&\leftarrow&
\left(\begin{array}{cc}
        \phantom{-}\cos\phi & \phantom{-}\sin\phi \\
        -\sin\phi & \phantom{-}\cos\phi
\end{array}\right)
{\mathbf y}_{1,n}
.
\label{T3}
\end{eqnarray}%
As a result the message is rotated by $\phi$
if the particle traveled via path 1.

\textbf{Simulation procedure.}
For each pair $(\alpha,\varphi)$,
$N=10000$ pairs of messengers (one for the photon, one for the ancilla)
are sent through the network (see Fig.~\ref{fig.3}) of processing units.
A messenger that appears on an output line of the network,
either exits via port 0 or via port 1, never via both ports simultaneously.
With each pair of messengers that emerges from the network,
the corresponding counter is incremented,
that is no events are being discarded.
In other words, we assume that the efficiency of the detectors is 100\%.
After all pairs have been processed, dividing the value
of one of the counters by $N$ yields the normalized frequency
for observing a pair (photon,ancilla) in the corresponding
output ports.

\section{Simulation results}

\begin{figure}[t]
\begin{center}
\includegraphics[width=8cm ]{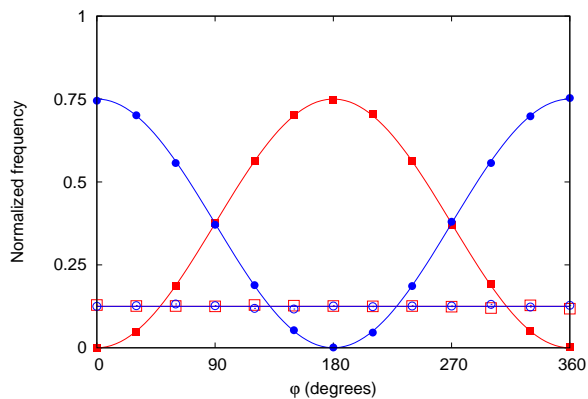}
\caption{%
The normalized frequency of observing a photon in
path 0 (squares) or 1 (circles) conditioned
on the observation of the ancilla in path 0 (open symbols)
or 1 (closed symbols), for the case
in which $\alpha=\pi/3$.
The solid lines are the prediction of quantum theory, see Eq.~(\ref{app15}) .
The number of emitted and detected events per $\varphi$ is 10000.
The DLM control parameter $\gamma=0.99$.
}%
\label{fig.4}
\end{center}
\end{figure}

In Fig.~\ref{fig.4}, we show results of the event-based simulation
of the quantum-controlled delayed-choice experiment for
a fixed value ($\alpha=\pi/3$) of the parameter that
determines the probability ($\sin^2\alpha$)
that the ancilla is in the state $|1\rangle$.
As the solid lines in Fig.~\ref{fig.4} are the predictions
of quantum theory, see Eq.~(\ref{app15}), it is clear that the event-based simulation
reproduces the results of quantum theory for this particular value of $\alpha$.

\begin{figure}[t]
\begin{center}
\includegraphics[width=8cm ]{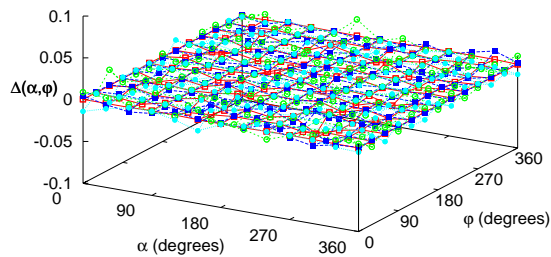}
\caption{%
Difference $\Delta(\alpha,\varphi)$ between the quantum theoretical result Eq.~(\ref{app15})
and the data obtained from an event-by-event simulation of
the quantum-circuit shown in Fig.~\ref{fig.3}.
The number of emitted and detected events per pair $(\alpha,\varphi)$ is 10000.
The DLM control parameter $\gamma=0.99$.
The differences fluctuate on the 1\% level.
Open squares: Photon detected in path 0, ancilla detected in path 0;
Closed squares: Photon detected in path 1, ancilla detected in path 0;
Open circles: Photon detected in path 0, ancilla detected in path 1;
Closed circles: Photon detected in path 1, ancilla detected in path 1.
Lines connecting markers are guide to the eye only.
}%
\label{fig.5}
\end{center}
\end{figure}

In Fig.~\ref{fig.5}, we plot
the difference between the event-based simulation results
and the prediction of quantum theory, given by Eq.~(\ref{app15}).
The differences are on the 1\% level, as it should be
on the basis of standard statistical arguments.
Therefore, we may conclude that the event-by-event approach
reproduces the statistical distributions of quantum theory
for the quantum-controlled delayed choice experiment.

\section{Discussion}

Instead of discussing our event-by-event simulation approach
for optical phenomena in full generality, in this paper we have
opted to explain in detail how the approach is applied to
a specific example, a quantum-controlled delayed-choice experiment~\cite{IONI11}.
We hope that this helps to understand the key feature of our approach,
namely that it builds, one-by-one,
the statistical distributions of quantum theory without knowing about the latter.

The successful simulation of the quantum-controlled delayed-choice experiment~\cite{IONI11}
adds one to the many examples for which the event-by-event simulation
method yields the correct statistical distributions.
Of course, the event-based approach, being free from concepts
such as particle-wave duality, does not suffer from
the conflicts with every-day logic that arise
in the quantum-theoretical description of the
delayed-choice experiment.
In particular, there is no need to invoke the thought
that in this experiment, the character of the photon
need to be changed in the past.

Finally, it should be noted that although
the discrete-event algorithm can be given an interpretation as a realistic cause-and-effect description that
is free of logical difficulties and reproduces the statistical results of quantum theory,
at present the lack of relevant data make it impossible to decide whether or not such algorithms are realized by Nature.
Only new, dedicated experiments that provide information beyond the statistics can teach us more about this intriguing question.

\bibliography{c:/d/papers/epr11,c:/d/papers/neutrons}   

\end{document}